%% file: Main.tex
\documentclass[5p,twocolumn,10pt,times]{elsarticle}
\usepackage{amsmath}
\usepackage{hyperref}
\usepackage{graphicx}
\usepackage{subfigure}
\usepackage{verbatim}
\usepackage{ulem}
\usepackage[ruled]{algorithm2e} 
\usepackage{algorithmic}
\usepackage{multirow}
\usepackage{caption}
\usepackage{amsmath}
\usepackage{ctable}
\biboptions{sort&compress}
\graphicspath{{images/}}
\usepackage{color}

\begin{document}

\begin{frontmatter}

\title{Field-Based Toolpath Generation for 3D Printing Continuous Fibre Reinforced Thermoplastic Composites}

\author[1,2]{Xiangjia Chen\corref{cor1}}
\author[3,4]{Guoxin Fang\corref{cor1}}
\author[1,5]{Wei-Hsin Liao}
\author[3]{Charlie C.L. Wang\corref{mycorrespondingauthor}}
\cortext[cor1]{Joint first authors}
\cortext[mycorrespondingauthor]{Corresponding author. E-mail: changling.wang@manchester.ac.uk}

\address[1]{Department of Mechanical and Automation Engineering, The Chinese University of Hong Kong, Shatin, Hong Kong, China}
\address[2]{Centre for Perceptual and Interactive Intelligence (CPII) Limited, Hong Kong, China}
\address[3]{Department of Mechanical, Aerospace and Civil Engineering, The University of Manchester, United Kingdom}
\address[4]{Faculty of Industrial Design Engineering, Delft University of Technology, The Netherlands}
\address[5]{Institute of Intelligent Design and Manufacturing, The Chinese University of Hong Kong, Shatin, Hong Kong, China}

\begin{abstract} 
We present a field-based method of toolpath generation for 3D printing continuous fibre reinforced thermoplastic composites. Our method employs the strong anisotropic material property of continuous fibres by generating toolpaths along the directions of tensile stresses in the critical regions. Moreover, the density of toolpath distribution is controlled in an adaptive way proportionally to the values of stresses. Specifically, a vector field is generated from the stress tensors under given loads and processed to have better compatibility between neighboring vectors. An optimal scalar field is computed later by making its gradients approximate the vector field. After that, isocurves of the scalar field are extracted to generate the toolpaths for continuous fibre reinforcement, which are also integrated with the boundary conformal toolpaths in user selected regions. The performance of our method has been verified on a variety of models in different loading conditions. Experimental tests are conducted on specimens by 3D printing \textit{continuous carbon fibres} (CCF) in a \textit{polylactic acid} (PLA) matrix. Compared to reinforcement by load-independent toolpaths, the specimens fabricated by our method show up to 71.4\% improvement on the mechanical strength in physical tests when using the same (or even slightly smaller) amount of continuous fibres.
\end{abstract}


\end{frontmatter}


\input{Text/Intro.tex} 
\input{Text/Sec2RelatedWork}
\input{Text/Sec3ExperimentalExploration}
\input{Text/Sec4FieldBasedToolpathGeneration}
\input{Text/Sec5Result}

\input{Text/Sec7Conclusion}

\section*{Acknowledgement}\vspace{-8pt}\noindent
The authors would like to thank the support from HKSAR RGC General Research Fund (GRF): CUHK/14202219. Guoxin Fang is supported in part by the China Scholarship Council.

\section*{References}
\bibliographystyle{elsarticle-num}
\bibliography{reference.bib}

\end{document}

%% file: Text/Intro.tex
\section{Introduction}\label{introduction}
As one of the most popular 3D printing techniques, fused filament fabrication has been widely employed in many applications because of its cost-effectiveness and large variety of material options for filament deposition~\cite{Gao15_CAD}. With growing demand to improve the mechanical strength of models fabricated by fused deposition, the reinforcement by integrating high-performance materials in fused filament fabrication has recently caught a lot of attention~\cite{Blok18_AM,Nekoda20_AdditManu}. 
In particular, the behavior of \textit{continuous carbon fibre} (CCF) reinforced thermoplastic composites has been well studied (ref.~\cite{Matsuzaki2016_SciReport, Brenken18_AM, Sanei20_JCS, Valvez20_PSI}). Models fabricated by CCF can provide excellent tensile strength along the axial direction of fibres while remaining lightweight. Moreover, a recent study showed that CCF is also a recyclable material~\cite{TIAN17_JCP}. 

At present, there are two major methods for 3D printing of continuous fibre reinforced polymers, including in-nozzle impregnation~\cite{Matsuzaki2016_SciReport} and out-of-nozzle impregnation~\cite{DICKSON2017146}. The benefit of out-of-nozzle impregnation is threefold. Firstly, dual extruders with one for thermoplastic materials (e.g., PLA, ABS, Nylon, etc.) as matrix and the other for continuous fibre are able to provide better control for stretching the continuous fibre during deposition. Secondly, the temperatures of continuous fibre and matrix resin can be separately controlled to achieve a better mechanical property. Lastly, the deposition of fibres can be adaptive and only applied in the critical regions that are necessary. Therefore, this technique is more widely used and even commercialized in the off-the-shelf 3D printers (e.g.,~\cite{Eiger}). On the other aspect, manufacturing parameters (temperature, layer thickness, platform temperature, etc.) have been studied to optimize the mechanical strength of CCF reinforcement~\cite{Tian16_composite, yang17_RPJ, heidari2019mechanical, shen2019study}. The out-of-nozzle impregnation strategy~\cite{DICKSON2017146} is employed in our research and physical experiment.

With the help of optimized paths for fibre alignment -- e.g., along the flow of stresses, the structures fabricated as \textit{continuous fibre reinforced thermoplastic composites} (CFRTPCs) can result in tensile strengths that are an order of magnitude higher than those of structures fabricated by preprag filaments with chopped fibres (ref.~\cite{Blok18_AM, Isobe18_MSE, Juan19_AM}). However, load-dependent toolpath generation for CCF reinforcement has not been well studied yet. In commercial CCF toolpath generators (e.g., Eiger software from MarkForged~\cite{Eiger}), hybrid of contour-parallel and zigzag-parallel toolpaths are employed, by which the performance of CCF reinforcement is not optimized. The recent work published in \cite{Li2020_CompositeB, Ting21_CompositesA} for generating load-dependent toolpaths was based on first selecting a critical region by topology optimization and then generating media-axis aligned toolpaths of CCF in selected regions. In contrast, we directly compute distributed toolpaths in the whole design space of input models by using the stress field obtained from \textit{finite element analysis} (FEA).

\begin{figure*}[t]
\centering 
\includegraphics[width=\linewidth]{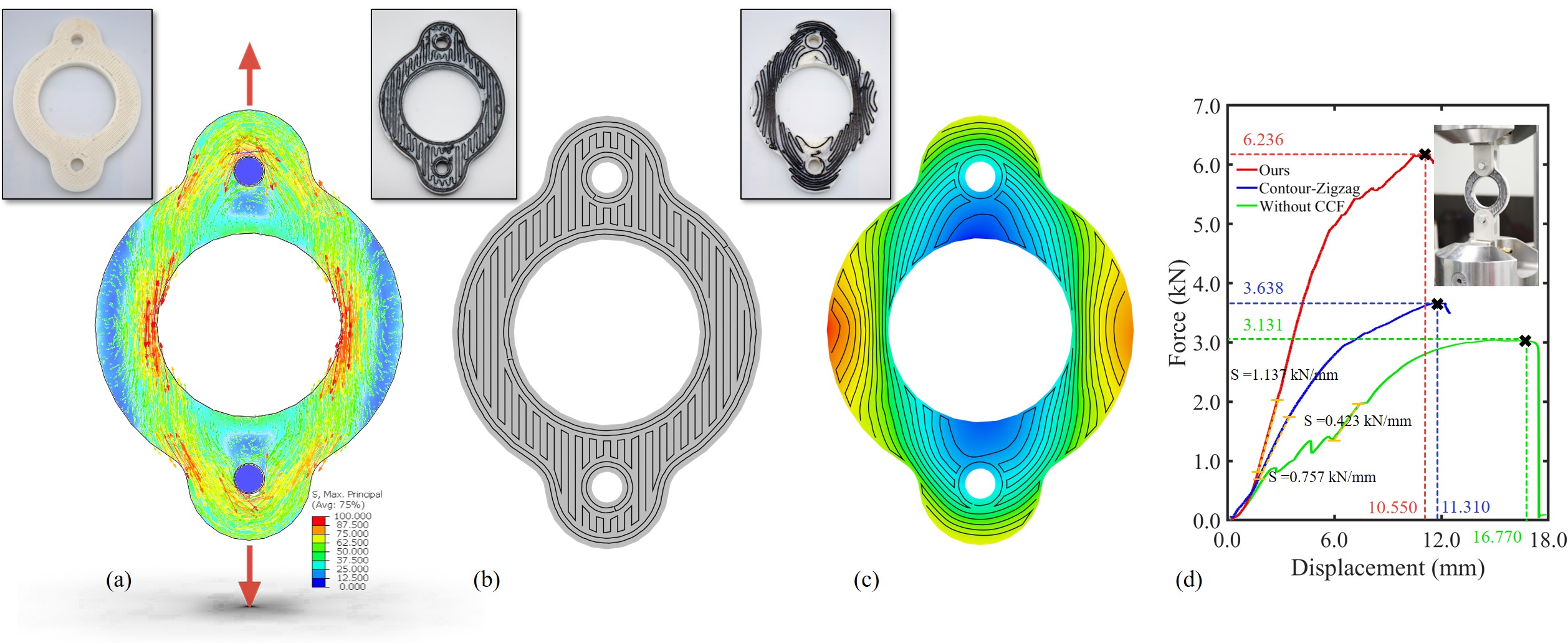}\vspace{-10pt}
\caption{Using the stress-field based toolpaths generated by our approach, the 3D printed CFRTPCs can result in much better mechanical strength in physical tests. (a) The field of principal stresses under a given load is generated by FEA. (b) The load-independent CCF toolpaths generated by off-the-shelf Eiger software \cite{Eiger} is employed as a benchmark to demonstrate the advantage of our approach. (c) Our method extracts isocurves from the governing scalar-field (colorful map) as CCF toolpaths, which is adaptive to the principal stresses in tension. 
(d) The force-displacement curves obtained from tensile tests by 3D printed specimens using the toolpaths in (b) and (c), where our method can increase the breaking force and the stiffness $S$ (as shown by the slope of the force-displacement curve) by 71.4\% and 50.2\% respectively. Note that, the total length of CCF filaments used in (b) and (c) are $35.895$~m and $34.658$~m -- i.e., ours are slightly shorter.}\label{fig:teaser}
\end{figure*}

In this paper, we propose a field-based computational method to generate CCF toolpaths along the directions of tensile stresses in the critical regions, which fully considers the strong anisotropic material property of continuous fibres for 3D printing CFRTPCs. The density of toolpath distribution generated by our method is adaptive -- i.e., denser in the region with higher stresses and sparser in the less critical regions (see Fig.~\ref{fig:teaser} for an example). By using well-controlled orientation and density of toolpaths in the fibre-layers between PLA layers, the breaking force of CFRTPCs fabricated by our method is $71.4\%$ higher than the specimens fabricated by using the load-independent CCF toolpaths (i.e., the zigzag parallel pattern with $45^{\circ}$ direction change between fibre layers). The PLA matrix in all specimens are all printed by using the zigzag parallel pattern -- again with $45^{\circ}$ change between layers. The enhancement in stiffness as the slope of force-displacement curve is $50.2\%$ by our method while using the similar length (or even slightly shorter) of CCF filaments in 3D printing. 

The major \textit{technical contribution} of our approach is a method to convert the stress-field obtained from FEA into a governing scalar-field to generate toolpaths for 3D printing CFRTPCs (Sections \ref{subsecVectorField} and \ref{subsecScalarField}). After obtaining the optimized governing field, we develop an algorithm to extract adaptive CCF toolpaths by controlling the minimally allowed distance between neighboring toolpaths (Section \ref{subsecAdaptivePath}). We also enhance the adaptive CCF toolpaths by merging the boundary conformal toolpaths that are added in user-specified boundary regions to reinforce the contact interface (Section \ref{subsecPathConnection}). Experimental tests on a variety of models in different loading conditions are conducted to verify the effectiveness of our approach and its advantage. 

The rest of this paper is organized as follows. After reviewing the literature in Section~\ref{secRelatedWork}, we conduct experimental exploration in Section \ref{secAnalysis} to derive the rules of CCF toolpaths for 3D printing CFRTPCs. Toolpath generation algorithms are detailed in Section \ref{secFieldBasedToolpath} as the major technical contributions of our paper. Both the computational and physical experiments are presented in Section \ref{secResults} together with the discussion on our method's limitations. 

%% file: Text/Sec2RelatedWork.tex
\section{Related Work}\label{secRelatedWork}
In this section, we review both the recent progress in 3D printing CCF reinforced composites and the related work on stress-oriented toolpath optimization in fused filament fabrication. A comprehensive review is beyond the scope of this paper, thus we only discuss the most relevant works below.

\subsection{Filament deposition-based fabrication of CFRTPCs}
Materials in continuous fibres such as carbon fibre, glass fibres and jute fibres have been used to fabricate CFRTPCs in complex shapes by 3D printing~\cite{Matsuzaki2016_SciReport,DICKSON2017146,Li21_AdditManu}. 
Among these materials, CCF is widely adopted because of its high stiffness and high tensile strength along the axial direction. Although the toolpath generation method presented in this paper can be applied to any continuous fibres, we employ CCF in our experimental tests as it is also used by the off-the-shelf 3D printers. 

Applying compaction by a roller is a widely used strategy for generating strong adhesion between CCF and matrix in \textit{automated fibre placement} (AFP) -- an industrial technology in composite material manufacturing \cite{Oromiehie2019_ATP}. Inspired by AFP, Omuro et al.~\cite{Omuro2017} installed a compaction roller on the printer head to `iron' fibres onto matrix in thermoplastics. Using a compaction roller in a printer head makes it hard to turn during printing; moreover, the position of fibre on the roller is unstable so that the quality of the 3D printing result is poor. This however motivated the development of out-of-nozzle 3D printing for CCF~\cite{Blok18_AM}, which is later employed in the off-the-shelf 3D printers with dual printer heads (e.g., Markforged~\cite{Eiger}) as stable extrusion of both CCF and melted thermoplastics can be realized. Specifically, the circular corner of the nozzle on a CCF printer head works as a compaction roller during the printing process to reform the CCF from a cylindrical filament into a planar ribbon. Instead of extruding by the motor, CCF filaments are in fact dragged out and then compacted by the circular corner. Therefore, the CCF ribbons are stretched during adhesion so as to result in better mechanical performance than those CFRTPCs fabricated by the in-nozzle impregnated filaments~\cite{Matsuzaki2016_SciReport, yang17_RPJ, heidari2019mechanical}. 

The influence of processing parameters on the mechanical strength of 3D printed CFRTPCs has been studied comprehensively in prior research (ref.~\cite{Tian16_composite,heidari2019mechanical,TIAN17_JCP,Wang21_AdditManu}). 
The `best' combination of parameters, such as the temperatures at the extruder and printing bed, the feed rate of the extruder and the movement speed of the printer head are usually obtained by empirical tests; which is not the focus of our work. Instead, we attempt to find the optimized toolpath w.r.t. the stress distribution under a given loading condition while keeping the aforementioned processing parameters unchanged. 

With the help of electronic microscopy images, researchers analyzed the failure mode of 3D printed CFRTPCs~\cite{Sanei20_JCS, Valvez20_PSI, Isobe18_MSE} and concluded that there are two major reasons for structural failure in CFRTPCs -- 1) delamination between fibres and matrix and 2) fracture of fibres. As the fracture of fibres is hard to improve, various studies have been conducted to improve the adhesion between CCF and the matrix in thermoplastic materials. A common method is to conduct post-processing treatment by applying pressure with high temperature -- e.g., improved bending modulus was reported in \cite{Omuro2017}. A recent study~\cite{Nekoda21_AdditManu} showed that an optimal result can be obtained with heating at $250^{\circ}\mathrm{C}$ under a pressure of 200 psi. In contrast, we aim to design toolpaths aligning with the tensile stress directions to reinforce 3D printed thermoplastic composites by utilizing the strong tensile strength of CCF meanwhile avoiding large transverse loads between continuous fibres. 

\subsection{Toolpath optimization for filament fused deposition}
For fused filament fabrication, the design of the toolpath takes a crucial role as the toolpath of filament alignment has a significant influence on the behavior of 3D printed models, in both surface quality~\cite{Shembekar19_JCICE, Etienne2019_SIG} and mechanical strength~\cite{Ahn02_RPJ, Vicente16_3DPAM, Fang20_TOG}. To improve mechanical strength, studies have been conducted to generate load-dependent toolpaths -- usually according to the stress distribution obtained by FEA.

Sugiyama \textit{et al.}~\cite{Sugiyama20_CST} generated CCF toolpaths by considering both the stress distribution and the geometry constraints, which showed good mechanical performance in physical tests. However, the toolpaths are manually designed and only presented on models with a simple bar shape. A thread of studies~\cite{Narasimha20_SciReport, Li2020_CompositeB, Ting21_CompositesA} came up with a design method of load-dependent toolpaths for fibre-reinforced 3D printing by using topology optimization to find a critical region for the placement of CCF. A truncation bias is introduced to exclude those `less' critical regions. 
A recently published work~\cite{Papapetrou20_compositeB} simultaneously conducted the topology optimization and the orientation optimization for fibres, which also excluded those less important regions from toolpath generation. On the other hand, we keep the geometry of the input model unchanged but control the density of CCF toolpaths in different regions to reflect the level of importance.

Methods have been developed whereby the stress field generated by FEA is directly incorporated into the algorithm to generate 3D printing toolpaths. Steuben et al.~\cite{steuben2016implicit} developed a method to generate toolpaths by the isocurves of the von Mises stress field as a scalar-field. The toolpaths generated from the von Mises stress field do not explicitly consider the direction of stresses, which makes it difficult to utilize the anisotropy of CCF's mechanical property to reinforce CFRTPCs. The other category of approaches employ vector field to generate toolpaths for reinforced 3D printing. Vectors of principal stresses are obtained by applying the Eigen vector decomposition on the stress tensors obtained from FEA. \textit{Principal stress lines} (PSL) are then extracted by tracing algorithms to be used as toolpaths (e.g.,~\cite{Eder21_JMP,tam2017additive}). However, the vector field directly formed by maximal principal stresses is neither smooth nor divergence-free. PSL obtained from such a field will not be smooth and can also have self-intersection. The study by Tian et al.~\cite{Tian2021_CMAME} computed a divergence-free vector field for the generation of intersection-free CCF toolpaths. However, all these methods which extract the CCF toolpath by particle tracing do not provide a good way to control the density / sparsity of toolpaths as the method presented in this paper. A graph-based searching algorithm presented in \cite{Xia20_AM} also has similar difficulties with density control. Our method is based on the pipeline of vector-field processing, scalar-field optimization and isocurve extraction~\cite{Fang20_TOG}. The new method is developed to process the vector-field by using \textit{minimal spanning tree} (MST) based search to assign orientations of vectors and re-generating vectors in the region with high `turbulence' with the help of minimizing a harmonic energy. Moreover, the computation of the scalar-field is modified to enable density control. Details can be found in Section \ref{secFieldBasedToolpath}.

%% file: Text/Sec3ExperimentalExploration.tex
\begin{figure}[t]
\includegraphics[width=\linewidth]{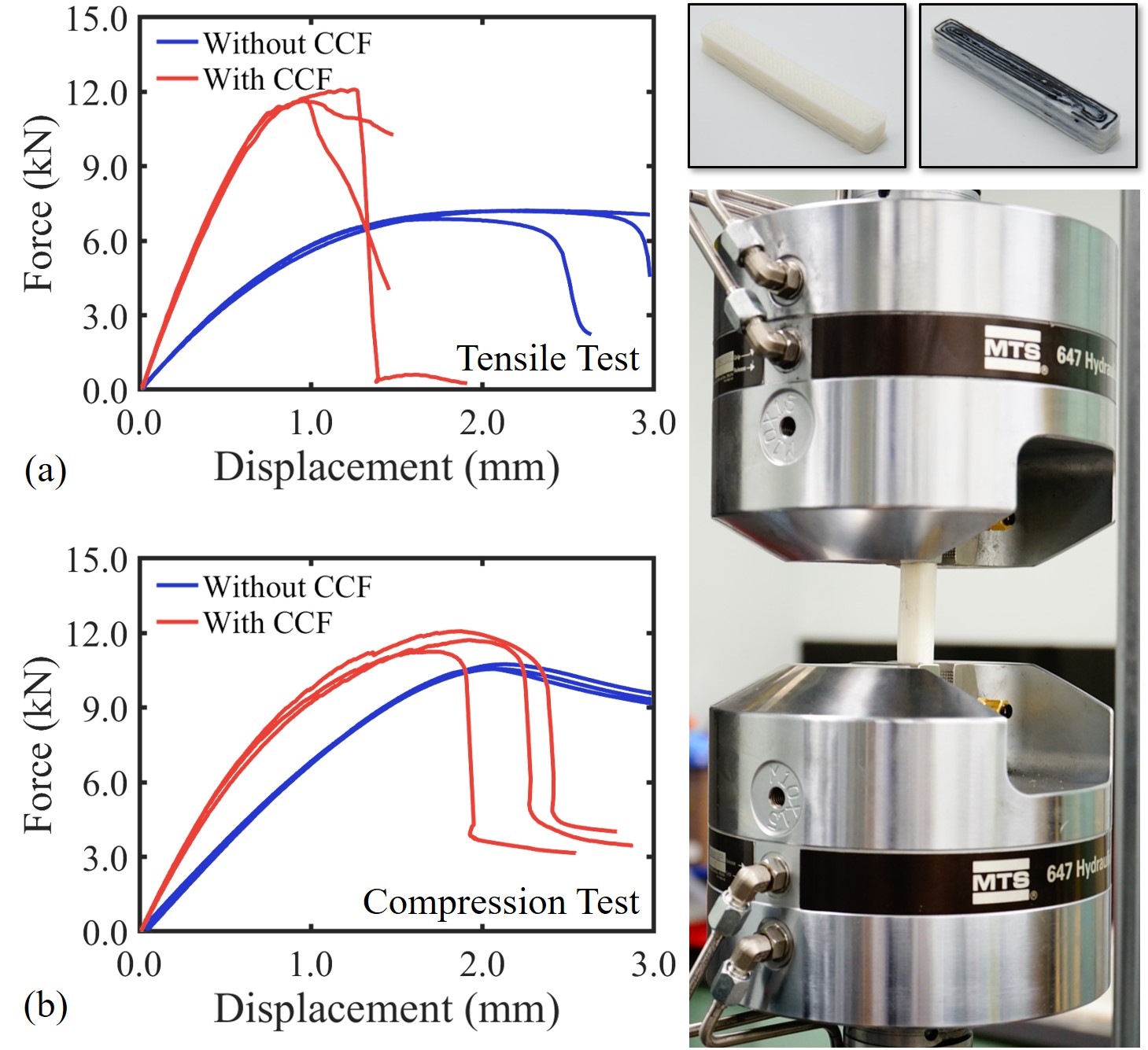}
\caption{Tensile and compression tests conducted on specimens with vs. without CCF reinforcement, where the layer of CCF is demonstrated in the image at the upper-right corner. The force-displacement curves are shown for (a) tensile and (b) compression tests respectively. The breaking force has been increased by $63.4\%$ (tensile) and $11.0\%$ (compression) after adding the CCF reinforcement layers. Three specimens are conducted for each test.}
\label{fig:CompressVSTensileTest}
\end{figure}

\begin{figure*}[t]
\includegraphics[width=\linewidth]{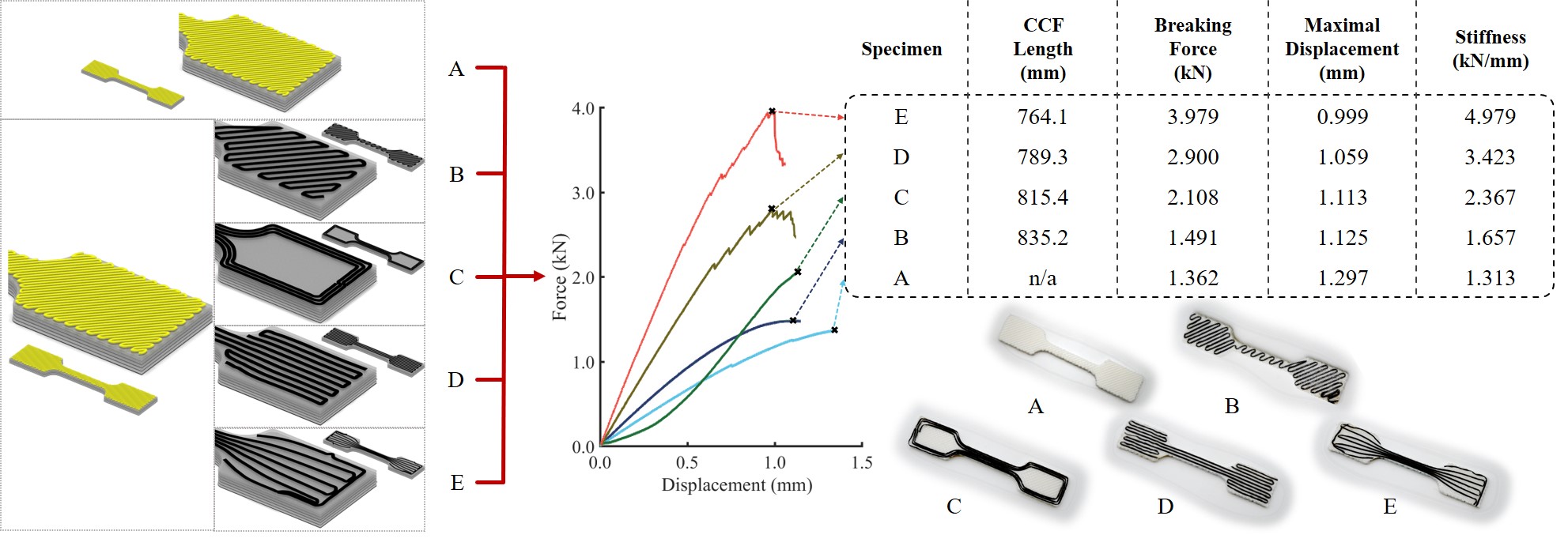}
\caption{Experimental results of tensile tests taken on models fabricated by using different schemes of CCF toolpaths -- (Model A) without CCF, (Model B) load-independent zigzag-parallel CCF toolpath, (Model C) load-independent contour-parallel CCF toolpath, (Model D) stress-oriented toolpath with uniform CCF density and (Model E) adaptive stress-oriented toolpaths.}
\label{fig:DiffPatternsTensileTest}
\end{figure*}

\section{Scheme for CCF Alignment: Experimental Exploration}\label{secAnalysis}
Continuous carbon fibres have been used in 3D printing to fabricate fibre reinforced composites because of the excellent mechanical property in high stiffness and tensile strength that can be achieved. However, previously published papers mainly presented the reinforcement in tensile tests. It is interesting to study and compare the effectiveness of reinforcement in both the tensile and the compression tests. There is a heuristic about aligning continuous fibres along the direction of principal stress lines. In this section, we will conduct experimental tests to explore the significance of this heuristic. Moreover, the influence of CCF's density will also be studied. 

Experimental tests are conducted on specimens in two different shapes, the shape of the tensile test bar and the shape of a cuboid. Note that the cuboid model is applied to study the effectiveness of reinforcement in tension vs. compression as its relatively large cross-sectional area can avoid buckling during compression tests. The force-displacement curves are obtained in experimental tests with 12 specimens fabricated for each of these two models. Therefore, each experiment is conducted on 3 specimens. Three curves are generated for each experiment as shown in Fig.~\ref{fig:CompressVSTensileTest}. The other tensile tests for exploring the strategy of CCF alignment are conducted on the specimens in the standard shape of the tensile bar (see Fig.~\ref{fig:DiffPatternsTensileTest}). Except the specimen without CCF (Model A), the toolpaths generated by different schemes are employed to fabricate models in different groups -- again 3 specimens per group. Model B and Model C were reinforced by using the zigzag-parallel and the contour-parallel toolpaths respectively, which were independent of the loading condition. The toolpaths employed in Model D are uniformly aligned along the directions of maximum stresses (in tension), and the adaptive density of stress-oriented toolpaths are employed in model E. Note that, the toolpaths in zigzag patterns are employed for the layers of resin matrix on all the specimens tested in this section. The same material, PLA, is used to fabricate the matrix of these models. The force-displacement curves are given in Fig.~\ref{fig:DiffPatternsTensileTest}) and the total lengths of CCF used for all models are also reported. As it is difficult to control the exact length of CCF used in a model, we adjusted the parameters of toolpath generation to use slightly shorter CCF filaments on models with stronger mechanical strength to ensure a fair comparison. 

Based on the results of experimental tests shown in Figs.~\ref{fig:CompressVSTensileTest} and \ref{fig:DiffPatternsTensileTest}, we can conclude the following observations.

\vspace{8pt} \noindent \textbf{Observation I:} \textit{Adding continuous carbon fibres as a very thin layer between the layers of resin matrix can enhance the mechanical strength of 3D printed models.} \vspace{8pt}

As can be found in all the tests conducted, adding intermediate CCF reinforcement layers with very thin thickness (e.g., $0.05$ mm in the specimens fabricated by our hardware system) can improve the mechanical strength in both the breaking force and the stiffness (as the slope of force-displacement curve shows). 

\vspace{8pt} \noindent \textbf{Observation II:} \textit{The reinforcement of adding CCF intermediate layers is more significant in the regions under tensile loads than those regions under compression.} \vspace{8pt}

It was found from the experiment shown in Fig.~\ref{fig:CompressVSTensileTest} that the mechanical strengthening caused by reinforcement was not significant under compression even if the CCF toolpaths are aligned along with the direction of maximum stresses. Therefore, only principal stresses in tension are considered in our scheme for CCF alignment.  

\vspace{8pt} \noindent \textbf{Observation III:} \textit{The reinforcement is more significant when aligning continuous fibres along the direction of large tensile stresses.} \vspace{8pt}

This observation is obtained by comparing the mechanical strength of Models B \& C with Models D \& E. Stronger mechanical strength is given on Models D \& E even when the amount of CCF used in these two models (i.e., $835.2$mm and $815.4$mm) is less than the Models B \& C (i.e., $789.3$mm and $764.1$mm). Based on this observation, our scheme for CCF alignment will control the direction of toolpaths according to the maximal principal stresses in tension. 

\begin{figure*}[t]
\includegraphics[width=\linewidth]{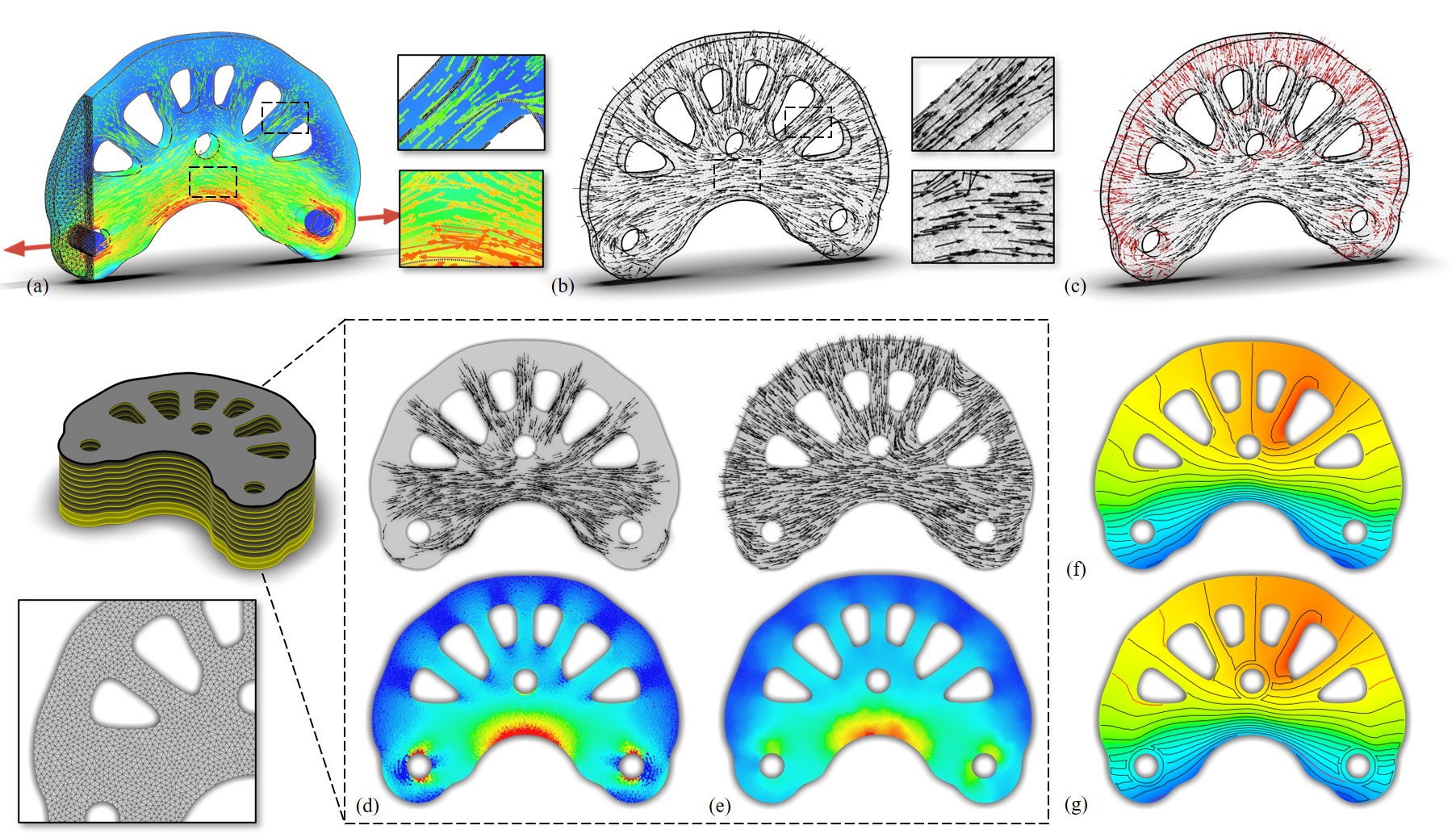}
\caption{The computational pipeline of our field-based toolpath generation algorithm. (a) The vectors of large principal stresses in tension are extracted in all tetrahedra of the input solid model. (b) Consistent orientations of the vectors are determined by propagation in an order determined by MST. (c) Vectors in `turbulent' regions (displayed in red) are identified and removed from the vector field $\mathbf{v}(\mathbf{x})$. (d) A slicing layer is partially spanned by the projected vectors of $\mathbf{v}(\mathbf{x})$. (e) The full vector field $\mathbf{u}(\mathbf{x})$ can be obtained by extrapolation / interpolation where the color map shows the distribution of weight $\sigma_f$ to control the density of CCF toolpaths. (f) The scalar field $s(\mathbf{x})$ is determined by letting its gradients $\nabla s(\mathbf{x}) \approx \mathbf{u}(\mathbf{x})$ and the isocurves of $s(\mathbf{x})$ are used as stress-oriented CCF toolpaths. (g) The final toolpaths are generated by merging the stress-oriented toolpaths with the boundary conformal toolpaths.}
\label{fig:AlgOverview}
\end{figure*}

\vspace{8pt} \noindent \textbf{Observation IV:} \textit{The reinforcement is further strengthened when aligning denser fibres in the regions with higher tensile stresses}. \vspace{8pt}

This observation is based on the comparison of Model D and Model E. It can be found that both the breaking force and the stiffness of Model E are further improved from Model D although less amount of CCF is used (i.e., $789.3$mm for Model D and $764.1$mm for Model E). 

Based on these experimental observations, we propose the following scheme of continuous fibre placement to aim at printing CFRTPCs with stronger mechanical strength.  

\vspace{8pt} \noindent \textbf{Scheme of CCF Alignment:} \textit{Reinforcement layers are fabricated by aligning CCF along the direction of tensile principal stresses in an adaptive  density that is proportional to the tensile stresses.} \vspace{8pt}

\noindent The algorithm developed according to this scheme will be presented in the following section, and the effectiveness of our algorithm will be verified on a variety of models under different loading conditions. 

%% file: Text/Sec4FieldBasedToolpathGeneration.tex
\section{Field-Based Toolpath Generation}\label{secFieldBasedToolpath}
This section presents the details of our field-based algorithm to generate CCF toolpaths for the reinforcement of 3D printed thermoplastic composites. The whole computational pipeline includes four major steps as illustrated in Fig.~\ref{fig:AlgOverview}. First of all, a vector field $\mathbf{v}(\mathbf{x})$ is computed in an input 3D solid $\mathcal{H}$ represented by a tetrahedral mesh $\mathcal{M}$, which is consistent with FEA. 
In the second step, the vector field is projected onto each layer $\mathcal{P}$ during slicing and converted into a scalar-field $s(\mathbf{x})$ ($\forall \mathbf{x} \in \mathcal{P} \cap \mathcal{H}$). 
In the third step, the CCF toolpaths are extracted from $s(\mathbf{x})$ by controlling the minimal distance between neighboring toolpaths.  
Lastly, the adaptive CCF toolpaths are merged with the boundary conformal toolpaths to form the final toolpath.

\subsection{Processing of vector field}\label{subsecVectorField}
Given a solid model $\mathcal{H}$ represented by a tetrahedral mesh $\mathcal{M}$, the stress distribution under a certain condition of loading can be computed by FEA and exported as a $3 \times 3$ stress tensors $\mathbf{\sigma}(e)$ for every element $e \in \mathcal{M}$. We can determine the principal stresses $[\sigma_1, \sigma_2, \sigma_3]$ (sorted by absolute values) and their corresponding directions $[\mathbf{\tau}_1, \mathbf{\tau}_2, \mathbf{\tau}_3]$ (as unit vectors) by applying the eigenvalue decomposition to $\mathbf{\sigma}(e)$. The vector $\mathbf{v}_e$ and the maximal tensile stress $\sigma_e$ for each element $e$ are defined by the following rules:
\begin{itemize}
    \item When $\sigma_1$ is positive (i.e., in tension), $\mathbf{v}_e = \mathbf{\tau}_1$ and $\sigma_e=\sigma_1$;
    
    \item If $\sigma_1$ is negative (i.e., in compression) but $\sigma_2$ is positive, $\mathbf{v}_e = \mathbf{\tau}_2$ and $\sigma_e=\sigma_2$ when $|\sigma_1 / \sigma_2| < \mu$ is satisfied\footnote{The threshold $\mu=3.0$ is employed in our implementation to avoid applying this rule to the regions that tension is too trivial than compression.};
    
    \item Otherwise, we consider there is no strong tensile direction in this region and $\mathbf{v}_e=\sigma_1$ and $\sigma_e=10^{-5}$ are assigned to avoid singularity in the following step of re-orientation.
\end{itemize}
Applying these results in a discrete field with vectors defined in the elements of $\mathcal{M}$ although there are some regions with undefined values. However, this vector field has issues of compatibility. Firstly, the direction of a principal stress is ambiguous -- i.e., either $\mathbf{\tau}_1$ or $-\mathbf{\tau}_1$ can be chosen by the decomposition algorithm as an eigenvector. As shown in Fig.\ref{fig:AlgOverview}(a), neighboring elements can have vectors with inverse orientations. Secondly, it is observed that the directions of principal stresses can have turbulent variation in some regions. Moreover, ambiguity is introduced in the region with isotropic stresses (i.e., $|\sigma_1| \approx |\sigma_2|$). Two methods, MST-based re-orientation and incompatibility removal, are introduced in our pipeline to process vectors to enhance the compatibility of neighboring toolpaths that are finally determined from the vector field. 

MST-based re-orientation is a method that has been widely used in computer graphics to generate consistent vectors as normals of point clouds for surface reconstruction \cite{Hoppe92}. We extend it from a surface domain into a volumetric domain here. Specifically, every tetrahedral element $e$ with a defined vector $\mathbf{v}_e$ is converted into a node of a graph $\mathcal{G}$. For two tetrahedra $e_i$ and $e_j$ that are face-neighbours to each other, an edge is constructed to link them in $\mathcal{G}$. Starting from a seed node, we can propagate the orientation of vectors by travelling on the graph. Without loss of the generality, when the tetrahedron $e_j$ is visited next after the tetrahedron $e_i$, the orientation of $\mathbf{v}_{e_j}$ will be flipped to $\mathbf{v}_{e_j}= -\mathbf{v}_{e_j}$ if $\mathbf{v}_{e_i} \cdot \mathbf{v}_{e_j} < 0$ (i.e., the orientations of vectors in $e_i$ and $e_j$ are inconsistent). As discussed in \cite{Hoppe92}, a na\"{i}ve breadth-first-traversal on $\mathcal{G}$ will mis-assign orientations in the regions with sharp change. To solve this problem, a weight 
\begin{equation}
    w_{i,j}= 1 - \left| \frac{\mathbf{v}_{e_i}}{\| \mathbf{v}_{e_i} \|} \cdot \frac{\mathbf{v}_{e_j}}{\| \mathbf{v}_{e_j} \|} \right|
\end{equation}
is given to every edge $(e_i,e_j)$. With the help of these weights, a better order of traversal is determined by computing MST on $\mathcal{G}$. The new order of node-visit can effectively avoid propagating orientations through the region with sharp change (i.e., edges with large weights). An example result of this step is shown in Fig.~\ref{fig:AlgOverview}(b), where a globally consistent vector field can be found although there is still some local incompatibility. 

A filter is applied to remove the vectors in locally incompatible regions, where the filtering algorithm is also implemented by using the connectivity of a tetrahedral mesh. For an element $e$, the set $\mathcal{N}_e$ of it neighbors is defined by all other elements that have a vertex, an edge or a face shared with $e$. We detect the region with incompatibility by finding all elements $e$ that give 
\begin{equation}\label{eqIncompatibleRemoval}
     \mathbf{v}_e \cdot \mathbf{v}_{e^*}  \leq  \eta \qquad (\forall e^* \in \mathcal{N}_e).
\end{equation}
See Fig.~\ref{fig:AlgOverview}(c) for an illustration displaying those incompatible vectors in red. Vectors of elements in incompatible regions are assigned as \textit{undefined} but the stress values $\sigma_e$ in these elements are retained. The threshold $\eta=0.5$ is determined by the statistical study of different examples (i.e., the histograms given in Fig.~\ref{fig:Histogram}). 

\subsection{Computing scalar field for toolpath generation}\label{subsecScalarField}
After obtaining a compatible vector field $\mathbf{v}(\mathbf{x})$ that is defined in the volumetric mesh $\mathcal{M}$ of the solid $\mathcal{H}$, we convert it into a scalar field $s(\mathbf{x})$ ($\forall \mathbf{x} \in \mathcal{P} \cap \mathcal{H}$) on a plane $\mathcal{P}$ during the slicing computation for 3D printing. A polygonal mesh $\mathcal{T}$ can be obtained by intersecting the tetrahedral mesh $\mathcal{M}$ with $\mathcal{P}$ (see the bottom-left of Fig.\ref{fig:AlgOverview}). For each polygonal face $f$ formed by intersecting a tetrahedron $e$ with the plane $\mathcal{P}$, the vector $\mathbf{v}_f$ can be obtained by projecting the vector $\mathbf{v}_e$ onto the $xy$-plane. The stress weight $\sigma_f$ of a face can also be derived from the tetrahedron $e$'s stress weight as $\sigma_f = \sigma_e$ ($\forall f \in e$). From Fig.\ref{fig:AlgOverview}(d), we can find that $\mathbf{v}_f$ is \textit{undefined} in some regions (i.e., those identified as incompatible and removed regions). Denoting the set of $f$'s neighboring faces by $\mathcal{N}_f$, the undefined values of $\mathbf{v}_f$ are determined by minimizing the following energy functions while smoothing the defined values / vectors of $\mathbf{v}_f$ and $\sigma_f$ in compatible regions.
\begin{equation}\label{eqSmoothEnergy1}
    E_{\sigma}=\sum_{f \in \mathcal{T}} \left( \frac{1}{|\mathcal{N}_f|}\sum_{f^* \in \mathcal{N}_f} (\sigma_f - \sigma_{f^*})^2\right)
\end{equation}
\begin{equation}\label{eqSmoothEnergy2}
    E_{\mathbf{v}}=\sum_{f \in \mathcal{T}} \left( \frac{1}{|\mathcal{N}_f|}\sum_{f^* \in \mathcal{N}_f} \| \mathbf{v}_f - \mathbf{v}_{f^*} \|^2\right)
\end{equation}
where $|\cdot|$ returns the number of elements in a set. To minimize these energy functions, a local Laplacian operator can be iteratively applied to update the values $\sigma_f$ (and the vectors $\mathbf{v}_f$) on all faces while fixing the value (and the vector) on the face with maximal $\sigma_f$ to impose a boundary condition for avoiding the trivial solution. In our implementation, the computation converges in 50 iterations. The processed vector field can then be obtained by
\begin{equation}\label{eqPlaneVectorField}
    \mathbf{u}_f= \sigma_f^p \mathbf{v}_f / \| \mathbf{v}_f\|
\end{equation}
with $p$ being a parameter to control the adaptation of CCF toolpath's density (more will be discussed in Section \ref{secResults}). Fig.\ref{fig:AlgOverview}(e) illustrates the results of processed fields for $\mathbf{v}_f$ and $\sigma_f$.

By defining the vectors in each face of $\mathcal{P}$, a piecewise linear scalar field $s(\mathbf{x})$ ($\forall \mathbf{x} \in \mathcal{P} \cap \mathcal{H}$) can be obtained by determining the field values on every node to minimize the difference between $\nabla s(\mathbf{x})$ and $\mathbf{u}_f$. That is
\begin{equation}\label{eqVector2Scalar}
    s(\mathbf{x}) = \arg \min \sum_{f \in \mathcal{T}} \| \nabla s(\mathbf{x}_f) - \mathbf{u}_f \|^2
\end{equation}
where $\mathbf{x}_f$ is the center of face $f$ and its scalar field value can be computed by piecewise linear function (see detail in~\cite{Fang20_TOG}). An example of the scalar field determined by Eq.(\ref{eqVector2Scalar}) can be found in Fig.\ref{fig:AlgOverview}(f). 

\subsection{Adaptive toolpath extraction}\label{subsecAdaptivePath}
Isocurves of the scalar field $s(\mathbf{x})$ ($\forall \mathbf{x} \in \mathcal{P} \cap \mathcal{H}$) are extracted as CCF toolpaths. When extracting too sparse isocurves to use as CCF toolpaths, 3D printed CFRTPCs cannot meet the expectation of reinforcement. When the isocurves are extracted too densely, overlapped CCF filaments will be generated during deposition. Overlapped CCF filaments will reduce the effectiveness of adhesion between CCF and the resin matrix and, therefore, also the mechanical strength of 3D printed CFRTPCs. Defining $W$ as the allowed minimal distance between neighboring CCF toolpaths -- a manufacturing parameter dependent on diameters of the nozzle and the CCF filament, we need to extract isocurves as dense as possible but with the minimal distance greater than $W$. For all examples conducted in our experiment, $1.0$mm as the width of CCF formed by deposition is employed for $W$.

We develop an algorithm to determine the isovalues for toolpath extraction by first increasing the number of isocurves $n$ to become dense enough and then incrementally reducing the value of $n$ until the requirement of minimal distance is satisfied.
\begin{itemize}
\item \textbf{Step 1:} Store the nodes with the minimal value $s_{\min}$ of the scalar field $s(\mathbf{x})$ in a set $\mathcal{S}$.

\item \textbf{Step 2:} For every node having the maximal field value $s_{\max}$, compute its approximate geodesic distance to $\mathcal{S}$ by the Dijkstra's algorithm and record the maximal value of these approximate geodesic distances as $D$.

\item \textbf{Step 3:} Let $n=\lceil D / W \rceil$ and extract the isocurves at \begin{equation}\label{eqIsocurves}
    s(\mathbf{x}) = s_{\min} + \frac{0.5 + i}{n} (s_{\max} - s_{\min}) \quad (i=0,1, \ldots,n-1)
\end{equation}
and compute the minimal distance $d$ between neighboring isocurves.

\item \textbf{Step 4:} If $d>W$, let $n= \lceil n * d / W \rceil$ and go back to Step 3. 

\item \textbf{Step 5:} Extract the isocurves by Eq.(\ref{eqIsocurves}) and compute the minimal distance $d$ between neighboring isocurves.

\item \textbf{Step 6:} If $d \leq W$, let $n=n-1$ and go back to Step 5.

\item \textbf{Step 7:} Output the isocurves as the CCF toolpaths.
\end{itemize}
After taking preparation in Steps 1-2 of this algorithm, Steps 3-4 increase the number of isocurves to an over-dense level. Then, Steps 5-6 incrementally reduce the density to ensure the required minimal distance $W$ between the neighboring CCF toolpaths. The toolpaths generated by isocurve extraction is denoted by $\mathcal{C}_{str}$. An example result can be found in Fig.\ref{fig:AlgOverview}(f).

\begin{figure}[t]
\includegraphics[width=\linewidth]{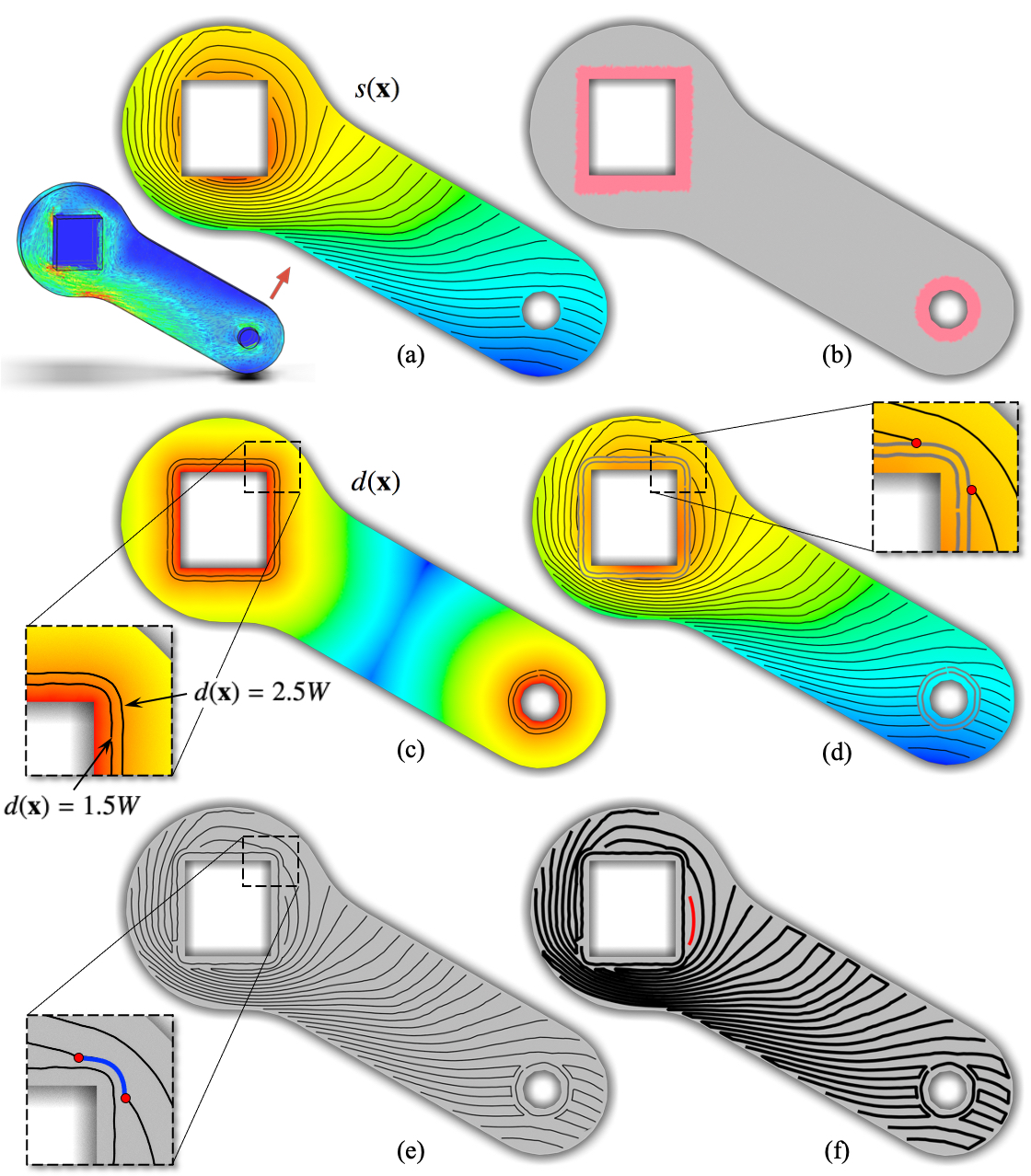}
\caption{Progressive results for the toolpath connection algorithm. (a) The toolpaths of $\mathcal{C}_{str}$ where the color map gives $s(\mathbf{x})$. (b) The boundary region selected by users as the contact interface to reinforce. (c) The boundary distance field $d(\mathbf{x})$ (color map) and the boundary conformal toolpaths $\mathcal{C}_{bnd}$. (d) The truncated $\mathcal{C}_{str}$ (color map is $s(\mathbf{x})$) with the curves in $\mathcal{C}_{bnd}$ displayed in gray, where the red dots in zoom-view are newly generated end points of $\mathcal{C}_{str}$. (e) The resultant toolpath after connecting the curves in $\mathcal{C}_{str}$ by part of the curve $d(\mathbf{x})=2.5W$ in $\mathcal{C}_{bnd}$. (f) The final CCF toolpaths, where the red segment will be excluded by considering the minimal manufacturable length.}
\label{fig:ToolpathConnectionMerging}
\end{figure}

\subsection{Toolpath connection}\label{subsecPathConnection}
One of the most common failures in CFRTPCs is the delamination between fibres and matrix (ref.~\cite{Sanei20_JCS, Valvez20_PSI, Isobe18_MSE}). To further enhance the mechanical bonding between 3D printed CCFs and the resin matrix, we further connect the isocurves obtained from different isovalues into continuously connected CCF toolpaths\footnote{Post-processing with high pressure and temperature~\cite{Omuro2017} can be applied to further enhance the bonding, which however is beyond the scope of this paper.}. Moreover, our method allows users to specify some boundary regions as interfaces of loading. Boundary conformal toolpaths that are generated in these regions and merged with the toolpaths $\mathcal{C}_{str}$ extracted from $s(\mathbf{x})$. 

\begin{table*}[t]
\centering 
\caption{Statistics of our computational pipeline}\label{tab:CompStatistic}
\footnotesize \vspace{-5pt}
\begin{tabular}{r|c|r||c|c||c|c|c|c|c||r}
\hline
        &   &   Element &  \multicolumn{2}{c||}{Time (sec.) for Processing 3D Vector Field}  &   &  \multicolumn{4}{c||}{Avg. Time (sec.) for Toolpath per Layer}   &   Total \\
\cline{4-5}  \cline{7-10}
{Model} & {Fig.} &  \#  & {MST-based Re-Ort.} & {Incompatible Removal} &  Layer \# & $\mathbf{u}_f$  & $s(\mathbf{x})$ & {Extraction} & {Connection} &  {Time (sec.)} \\
\hline
\hline
Loop &  \ref{fig:teaser}  &  32,473 &  17.684 &  0.810 &  12 &  0.200 &  0.030 & 8.210 & 0.591 & 130.862  \\
Shell\_Tensile &  \ref{fig:AlgOverview} & 112,080  &  400.683 &  3.262 &  12 &  1.093 &  0.096 & 24.077 & 1.225 & 758.140  \\
Wrench &  \ref{fig:ToolpathConnectionMerging} & 14,002  &  2.883 &  0.329 &  12 &  0.108 &  0.016 &  3.414 &0.342 & 56.642  \\
Piston &  \ref{fig:Piston} &  85,558 &  215.88 &  2.501 &  66 &  0.246 &  0.031 & 6.035 & 0.462 & 665.953  \\
Shell\_Compress & \ref{fig:ShellCompress}  & 112,080  &  414.114 &  3.117 &  12 &  0.964 &  0.162 & 16.732 & 0.935 & 676.000  \\
\hline
\end{tabular}
\end{table*}

Specifically, the following algorithm is applied to generate the CCF toolpaths as the outcome of our approach. The algorithm is described with the help of the example shown in Fig.~\ref{fig:ToolpathConnectionMerging}(a). 
\begin{itemize}
\item \textbf{Step 1:} A boundary distance field $d(\mathbf{x})$ is generated by the heat diffusion method \cite{Crane2017HMD} on the mesh $\mathcal{T}$, where only the user-specified boundaries (e.g., the region shown in Fig.~\ref{fig:ToolpathConnectionMerging}(b)) are employed as the heat source. Two isocurves $d(\mathbf{x})=1.5W$, $2.5W$ are extracted to serve as the boundary conformal toolpaths $\mathcal{C}_{bnd}$ (see Fig.~\ref{fig:ToolpathConnectionMerging}(c)) with $W$ being the width of CCF formed by deposition.

\item \textbf{Step 2:} For all curves in $\mathcal{C}_{str}$, the portion of curves with $d(\mathbf{x})<2.5W$ are truncated and removed from $\mathcal{C}_{str}$. This truncation generates new endpoints on the curve $d(\mathbf{x})=2.5W$ of $\mathcal{C}_{str}$ (see Fig.~\ref{fig:ToolpathConnectionMerging}(d) for placing the truncated $\mathcal{C}_{str}$ and  curves of $\mathcal{C}_{bnd}$ that are placed together, where the red dots in zoom-view are newly generated end points). Note that the toolpath with $d(\mathbf{x})=0.5W$ is neglected as it is too close to the boundary to have stable CCF deposition.

\item \textbf{Step 3:} For each newly generated endpoint in Step 2, connecting it to its nearby new endpoint along the curve $d(\mathbf{x})=2.5W$ (e.g., the part displayed in blue in the zoom-view of Fig.~\ref{fig:ToolpathConnectionMerging}(e)). After connecting all new endpoints, removing the originally extracted curve $d(\mathbf{x})=2.5W$ in $\mathcal{C}_{bnd}$. The result after this step is illustrated in Fig.~\ref{fig:ToolpathConnectionMerging}(e).

\item \textbf{Step 4:} For any remaining endpoints of curves in $\mathcal{C}_{str}$, we search and connect them to another endpoint to form a continuous toolpath when the distance between these two endpoints is less than $2W$. 
\end{itemize}
An example of the final result generated by the above algorithm is given in Fig.~\ref{fig:ToolpathConnectionMerging}(f). Moreover, considering the fibre-cutting mechanism used in the printer head, toolpaths with lengthes that are tool short (e.g., less than $42\mathrm{mm}$ in our hardware setup) are removed from the final CCF toolpaths. 

%% file: Text/Sec5Result.tex
\section{Results and Discussions}\label{secResults}

\subsection{Computational Results}
We implemented our computational framework in C++. The source code and the datasets of this work will be made available to the public. All the computational experiments were obtained on a desktop PC with an Intel(R) Core TM i7-9750H CPU (6 cores @ 2.6GHz) + 32GB RAM running Windows 10. The numerical library Eigen~\cite{eigenweb} is employed as the solver of linear equations. 

\begin{figure}[t]
\includegraphics[width=\linewidth]{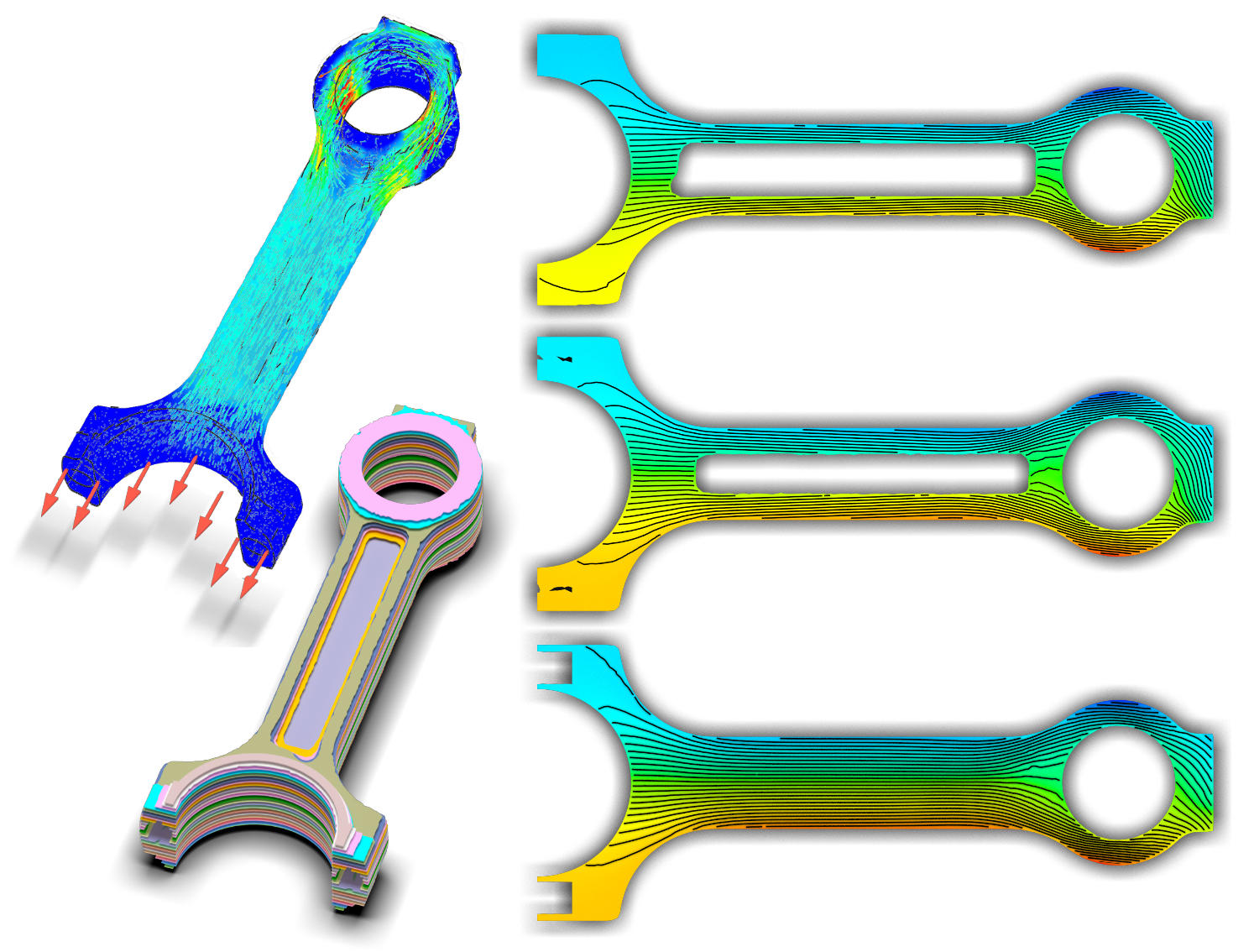}
\caption{A \textit{Piston} model with different contour shapes in different layers -- different CCF toolpaths are generated on different layers.
}\label{fig:Piston}
\end{figure}
\begin{figure}
\includegraphics[width=\linewidth]{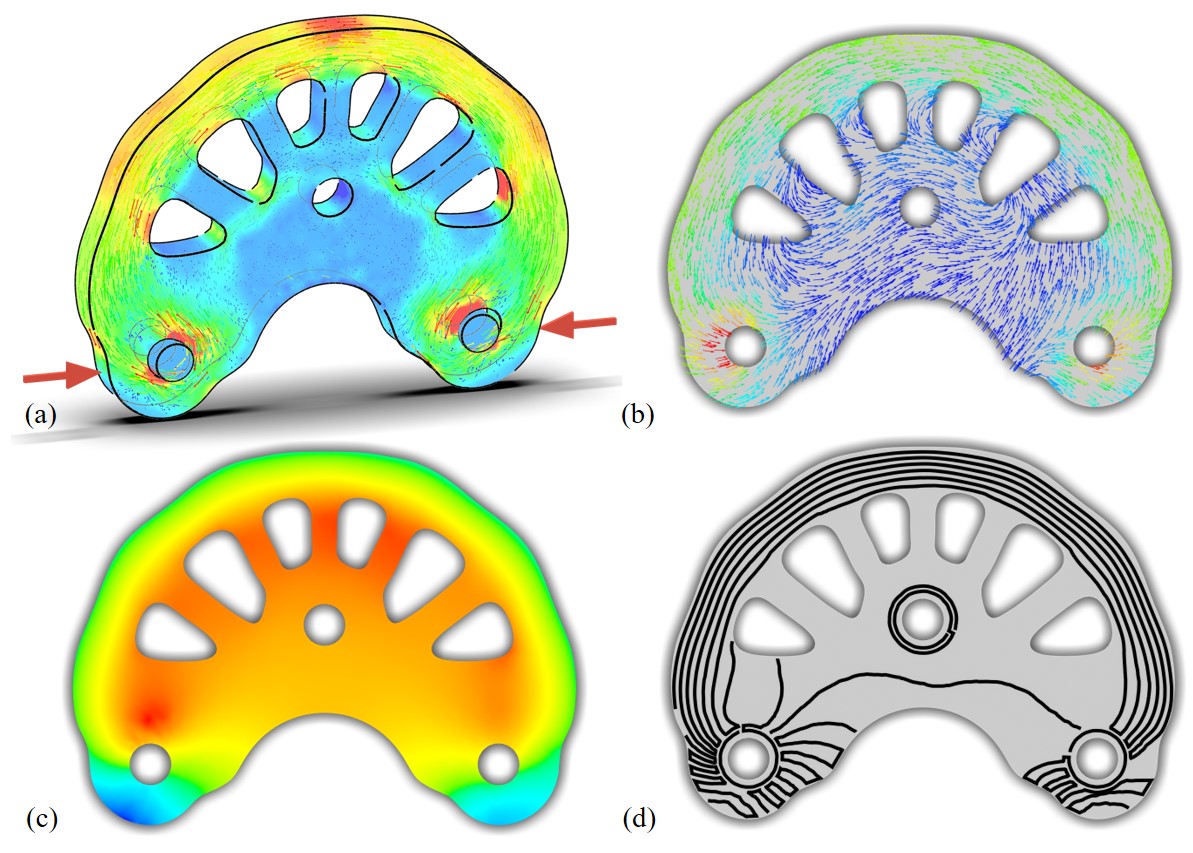}
\caption{When changing the loading condition from tensile into compression (a), the vector field $\mathbf{u}(\mathbf{x})$ (b) and the scalar field $s(\mathbf{x})$ (c) which are different from Fig.~\ref{fig:AlgOverview}(e) and Fig.~\ref{fig:AlgOverview}(f) will be generated. This also leads to a different CCF toolpath (d). 
}
\label{fig:ShellCompress}
\end{figure}

We have tested our approach on a variety of models with different loading conditions. The first example is the~\textit{Loop} model shown in Fig.~\ref{fig:teaser} with the tensile loading applied on the holes located on the top and bottom of the model. The second model is the \textit{Shell\_Tensile} model with complex cross-section as shown in Fig.~\ref{fig:AlgOverview}. The third is a \textit{Wrench} model as given in Fig.~\ref{fig:ToolpathConnectionMerging}, and the forth is a \textit{Piston} model having different contour shapes in different layers for 3D printing (see Fig.~\ref{fig:Piston}). The last example shows that different toolpaths are generated when changing the loading condition of the Shell model from tensile into compression (see the \textit{Shell\_Compress} model shown in Fig.~\ref{fig:ShellCompress}). 

The computational statistics are given in Table \ref{tab:CompStatistic}. It was found that the computation of all examples can be finished in less than 13 minutes. The major bottleneck of our computation is the toolpath extraction step -- it takes around $93.8\%$ to $97.3\%$ of the toolpath generation time per layer. The MST-based vector re-orientation step is also relatively time-consuming when the number of tetrahedra is large.

\begin{figure*}[t]
\includegraphics[width=\linewidth]{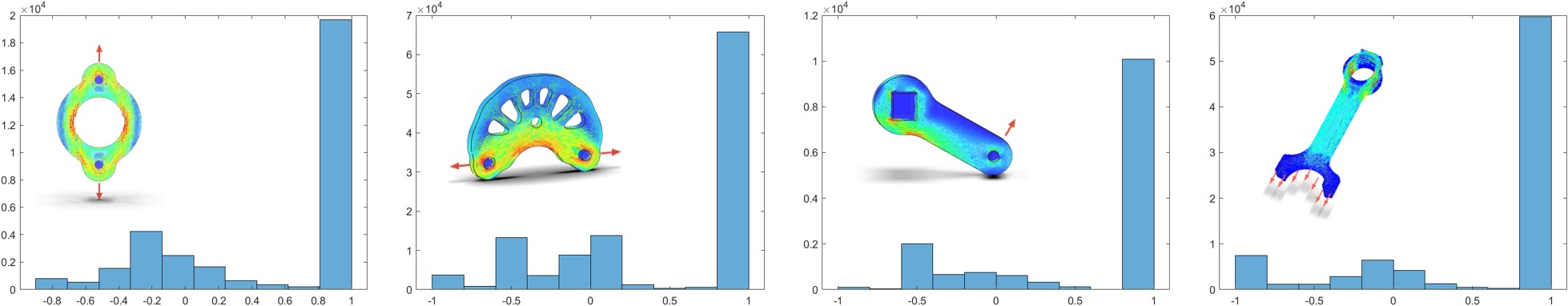}
\caption{$\forall (e,e^*) \in \Gamma$ with $\Gamma$ denoting all pairs of neighboring elements in a model, histograms of $(\mathbf{v}_e \cdot \mathbf{v}_{e^*})$ are generated for the vector fields after applying the MST-based reorientation. From left to right, the histograms of four models -- Loop, Shell\_Tensile, Wrench, Piston -- are shown. Using $\eta=0.5$ as the threshold in Eq.(\ref{eqIncompatibleRemoval}) can effectively segment the pairs into compatible vs. incompatible regions.
}
\label{fig:Histogram}
\end{figure*}
\begin{figure*}[t]
\includegraphics[width=\linewidth]{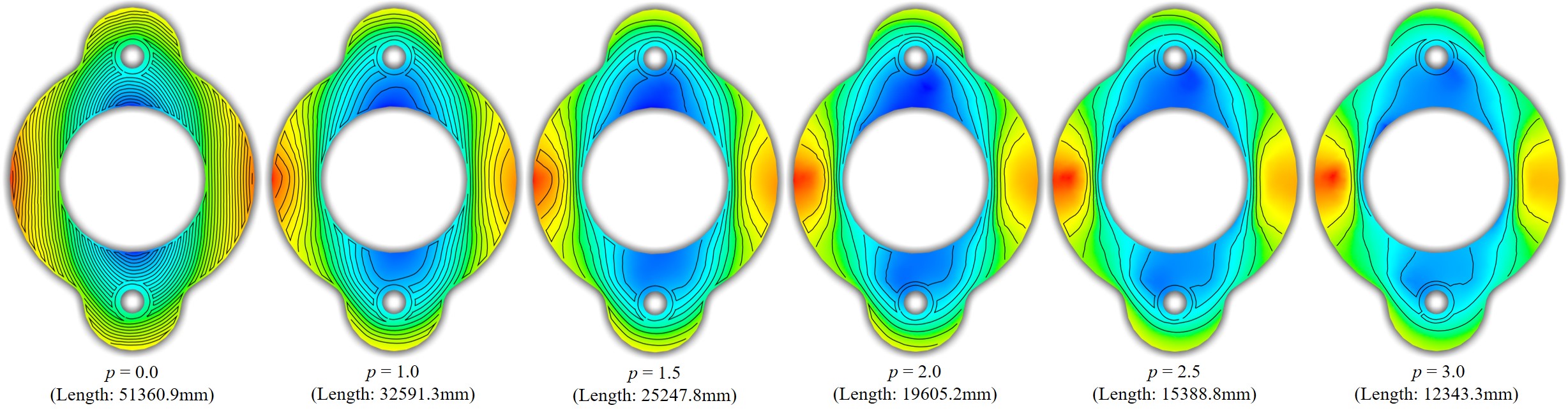}\\
\vspace{-20pt}
\caption{Density control can be realized by changing the value of parameter $p$ employed in Eq.(\ref{eqPlaneVectorField}). Most results with adaptive CCF toolpaths shown in this paper use $p=1.0$ while the result with uniform CCF toolpaths can be generated by $p=0.0$. Different lengths of the CCF toolpaths are also reported in this figure when using different $p$ values.}\label{fig:DensityControl}
\end{figure*}

We now study the selection of a few parameters employed in our algorithm. First of all, the threshold $\eta$ used in Eq.(\ref{eqIncompatibleRemoval}) for the incompatibility removal is studied. Histograms are generated for the dot products on all pairs of neighboring elements' vectors before applying the incompatible removal. The resultant histograms for different models are shown in Fig.\ref{fig:Histogram}. It can be found that the regions of compatible vectors can be effectively separated from the incompatible regions by using $\eta=0.5$. 

Another interesting study is about the parameter $p$ used in Eq.(\ref{eqPlaneVectorField}) for controlling the contrast of CCF toolpath's density. When using larger values for $p$, more significant variation of the CCF toolpaths' density will be generated (see Fig.~\ref{fig:DensityControl} for examples). For a special case, uniform CCF toolpaths will be generated when $p=0.0$ is employed. We leave the value of $p$ as a parameter to be tuned by users and $p=1.0$ is used to generate adaptive CCF toolpaths in all examples presented in this paper. 

\begin{figure}[t]
\includegraphics[width=\linewidth]{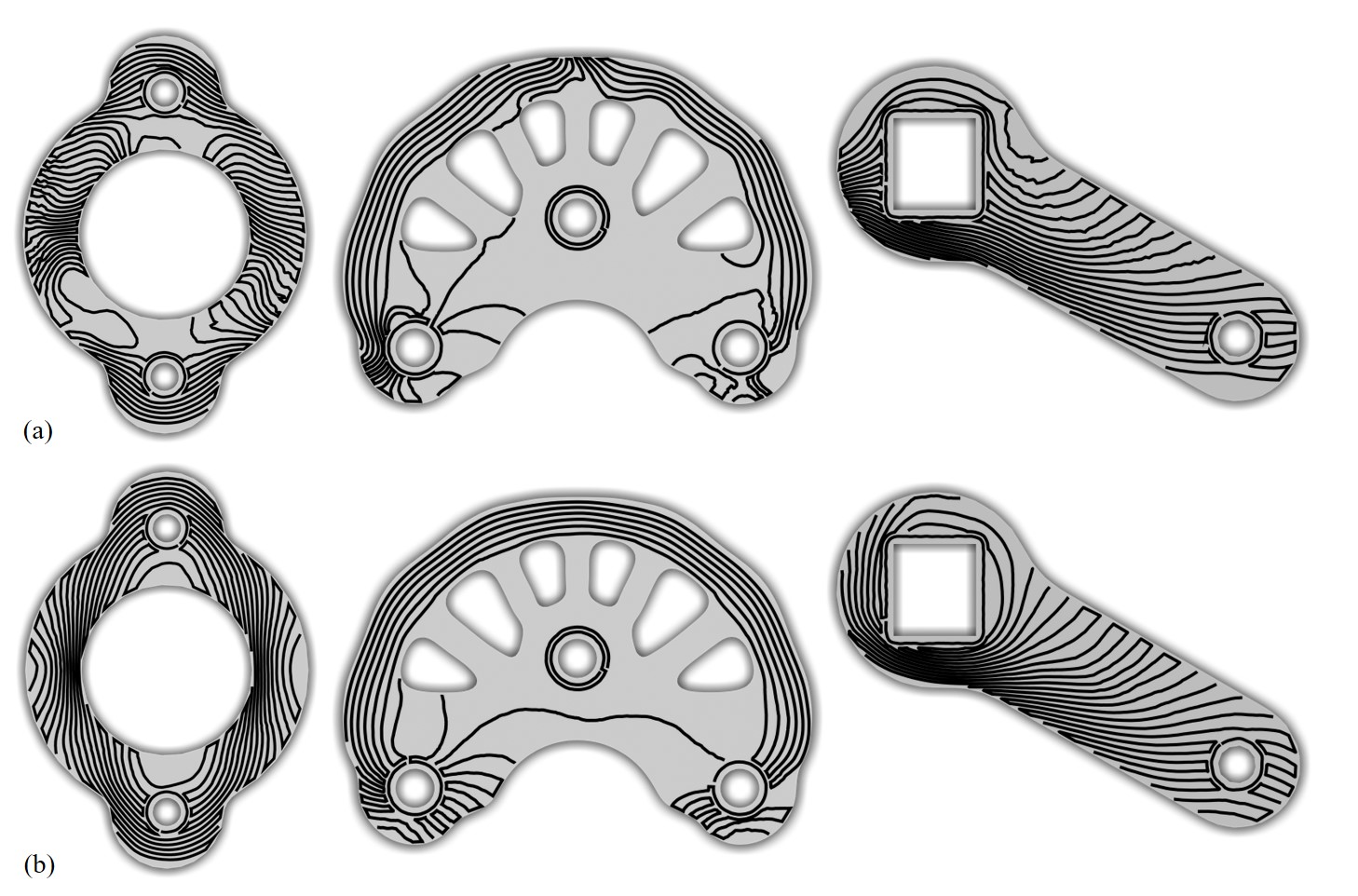}\\
\vspace{-20pt}
\caption{The comparison to show the effectiveness of field processing by MST-based reorientation and incompatible removal -- (a) the results without these steps of field processing vs. (b) the resultant toolpaths generated from processed fields.}
\label{fig:SmoothRes}
\end{figure}

Lastly, we study the effectiveness of MST-based reorientation and incompatibility removal for vector field. The results of direct computing from the 3D vector fields without these processing steps are given in Fig.\ref{fig:SmoothRes}(a). Unwanted turnings and oscillation can be observed in the results of all three examples. After using MST-based reorientation to generate vectors with consistent orientation and removing those incompatible regions, smooth CCF toolpaths can be generated by our approach (see Fig.\ref{fig:SmoothRes}(b)). 

\vspace{-10pt}

\subsection{Hardware and parameters of manufacturing}
\begin{figure}
\centering
\includegraphics[width=\linewidth]{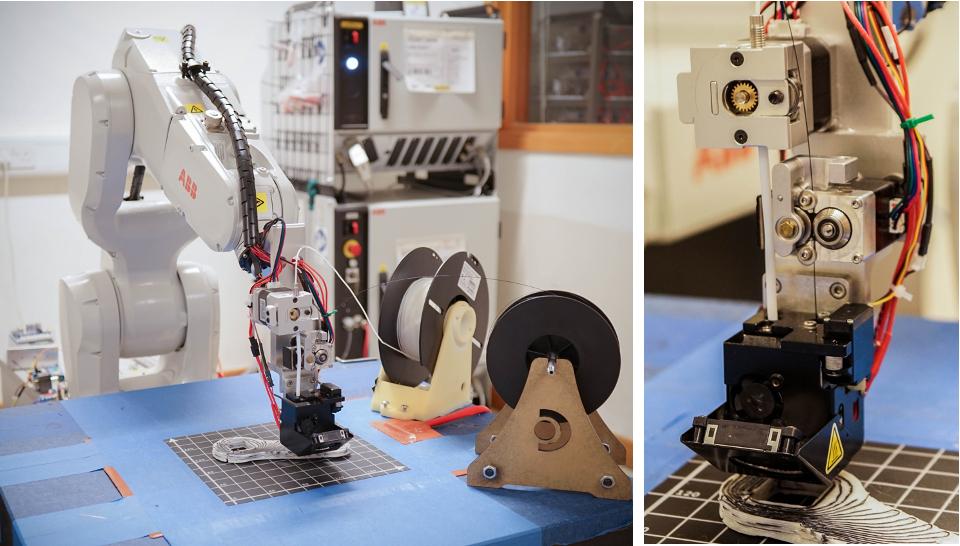}
\caption{The hardware setup for fabricating specimens used in our physical experiment to test the mechanical properties when different toolpaths are applied.}
\label{fig:Hardware}
\end{figure}

Models presented in our paper are fabricated by an IRB 1200 6-DOF robot arm with a dual extruder installed on the end-effector (see Fig.\ref{fig:Hardware}). The working space of this hardware setup is $300 \mathrm{mm} \times 300 \mathrm{mm} \times 300 \mathrm{mm}$. The CCF printer head and the PLA printer head of the dual extruder are installed in parallel along the z-axis. With individual control systems, the dual extruder enables 3D printing at different rates and temperatures for CCF and PLA filaments. The nozzle for PLA filament is similar to the nozzle commonly used on a conventional 3D printer while the other nozzle for CCF is equipped with a large, rounded corner at its end to compress the fibre filament onto the resin matrix during printing. Moreover, the CCF printer head also contains a cutter to chop the CCF filament when the deposition goes to the end of a toolpath. However, there is a distance between the cutter and the nozzle -- $42 \mathrm{mm}$ for the CCF printer head we used. As a result, any toolpath with length less than this distance cannot be realized for CCF deposition. After already trying to connect the toolpaths into a continuous one (i.e., Step 4 of the algorithm presented in Section \ref{subsecPathConnection}), we simply removed those with length less than $42\mathrm{mm}$.

To enable the synchronous motion between the extruder and the robot arm, we employed an Arduino board GT2560 for data communication, which supports the dual extruder and can provide extra I/O ports to control the cutter of the CCF printer head.
As the two nozzles of CCF and PLA are installed in parallel, we controlled the motor at the last joint of the robot arm to tilt at a small angle (i.e., $\pm 15^\circ$) at its end-effector to avoid collision between the already printed part and the nozzle not under working.

The parameters for 3D printing CFRTPCs in our experiment are listed in Table \ref{tabPrintingParameters}, where these parameters are employed according to prior studies \cite{Blok18_AM, Valvez20_PSI, Sanei20_JCS}. Our experiment also verifies that specimens of CFRTPCs in good quality can be successfully fabricated by using these parameters. We kept these parameters unchanged for fabricating all examples shown in this paper.

\begin{table*}[t]
\centering \footnotesize 
\caption{Statistics of experimental tests on specimens}\label{tab:PhysicalStatistic}\vspace{-5pt}
\begin{tabular}{r|c|r||c|c|c|c||c|c}
\hline
     &   &  Dimension & CCF Toolpath &  \multicolumn{2}{c|}{Material Usage -- Filament Length} &   Fabrication & \multicolumn{2}{c}{Mechanical Strength} \\
\cline{5-6} \cline{8-9}
 Model & Figure & (Unit: mm) &  Strategy & CCF (meter) & PLA (meter) &  Time (min.) & Breaking Force & Slope of Curve$^\ddag$\\
\hline
\hline
    &  \ref{fig:teaser}(a) \& (d) &  L: 140.00  & n/a        &  -  &  &  96  &  3.131~kN  &  0.423\\
Loop   &  \ref{fig:teaser}(b) \& (d) & W: 100.00 & Contour-Zigzag$^\dag$  &  35.895  &  60.619  &  230  &  3.638~kN &  0.757 \\
    & \ref{fig:teaser}(c) \& (d) &  H: 10.00   & Ours       &  34.658  &   &  221  &  6.236~kN &  1.137 \\
\hline
    &  &  L: 146.16  & n/a      &  -  &    &  85  &  0.903~kN  &  0.074\\
  Wrench      &  \ref{fig:PhysicalTestRes}(a) &W: 105.98 & Contour-Zigzag &  26.634  &  59.342  &  184  &  1.304~kN  &  0.083 \\
   &  &H: 10.00   & Ours       &  25.670  &    &  172  &  1.476~kN &  0.098 \\
\hline
    &  &  L: 144.57  & n/a    &  -  &    &  145  &  2.662~kN  & 0.729 \\
 Shell\_Tensile       &  \ref{fig:PhysicalTestRes}(b) &  W: 107.11 & Contour-Zigzag &  36.998 &  93.368  &  288  &  3.837~kN  & 1.087  \\
        &  &H: 10.00   & Ours    &  24.859  &    &  223  &  5.643~kN &  1.406 \\
\hline
    &  &  L: 144.57  &  n/a  &  -  &    &  145  &  3.419~kN  &  0.724\\
  Shell\_Compress               &  \ref{fig:PhysicalTestRes}(c) &W: 107.11 & Contour-Zigzag &  36.998   &  93.368  &  228  & 4.519~kN &  0.997 \\
                &  &H: 10.00   &  Ours  &  35.635   &    &  215  &  5.333~kN  &  1.062 \\
\hline
\end{tabular}
\begin{flushleft}
$^\dag$~Contour-Zigzag means load-independent CCF toolpaths that are generated in a way similar to the  off-the-shelf Eiger software \cite{Eiger}, where the boundary conformal and the zigzag parallel toolpaths are applied to the boundary and the interior regions respectively.\\
$^\ddag$~The slopes of force-displacement curves (Unit:~kN/mm) are reported to evaluate the stiffness of 3D printed specimens -- the bigger the higher stiffness is presented.
\end{flushleft}
\end{table*}

\begin{table}[t]
\centering \footnotesize 
\caption{Parameters employed in 3D printing process.}\vspace{-5pt}\label{tabPrintingParameters}
\begin{tabular}{c||c|c}
\hline
            & eSUN 1.75 mm & F-FG-0005 \\
Filament Type   & PLA+ & Carbon Fibre CFF Spool \\
\hline
Filament Diameter & 0.8 mm  & 1.0 mm  \\
\hline
\hline
Printing Speed  & 12 mm/s & 5 mm/s \\
Temperature & 200 $^\circ$C \par & 250 $^\circ$C \par \\
Layer Height & 0.8 mm & 0.05 mm \\
Line Width  & 0.8 mm & 1.2 mm \\
\hline
\end{tabular}
\end{table}

\begin{figure}
\centering
\includegraphics[width=\linewidth]{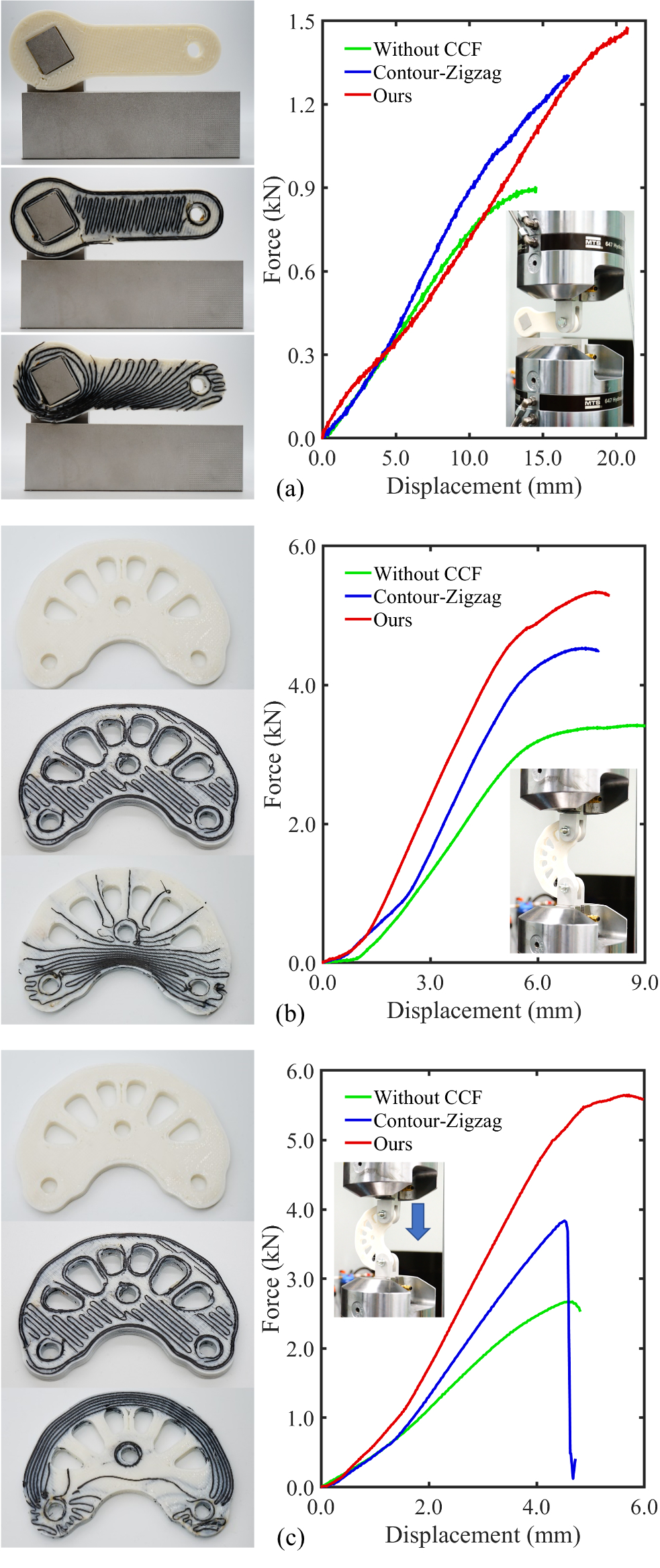}
\caption{The 3D printed specimens with and without CCF reinforcement and the resultant force-displacement curves in mechanical tests, where the results fabricated by load-independent CCF toolpaths as contour-zigzag are compared with the models reinforced by using our stress-oriented adaptive CCF toolpaths.}\label{fig:PhysicalTestRes}
\end{figure}

\subsection{Physical experiment}
The effectiveness of our CCF toolpaths is verified on four models, where each is fabricated by using both our CCF toolpaths and the contour-zigzag CCF toolpaths (i.e., the load-independent ones which are similar to the results of the off-the-shelf Eiger software \cite{Eiger}). The mechanical strength on both specimens were compared with the 3D printed model without CCF reinforcement. The mechanical tests were conducted on a Landmark Servohydraulic Test System with specially designed fixtures made by aluminium alloy. The resultant force-displacement curves are shown in Figs.\ref{fig:teaser} and \ref{fig:PhysicalTestRes}, and the corresponding data of fabrication and mechanical strength are listed in Table \ref{tab:PhysicalStatistic}.

As there are additional layers of CCF between every two layers of PLA, the fabrication time of CFRTPCs is more than double of the models 3D printed by PLA only. When using filaments of the same length (or slightly shorter), the CFRTPCs fabricated by our toolpaths demonstrate 99.2\%~(Loop), 63.5\%~(Wrench), 112.0\%~(Shell\_Tensile) and 56.0\%~(Shell\_Compress) strength enhancement in comparison to the contour-zigzag toolpaths which only give 16.2\%~(Loop), 44.4\%~(Wrench) 44.1\%~(Shell\_Tensile) and 32.2\%~(Shell\_Compress) improvement in the breaking force. Moreover, the slopes of the force-displacement curves obtained from our method are in the range of a $6\%$ to $50\%$ improvement to the load-independent contour-zigzag toolpaths. All these results have also been demonstrated in our supplementary video that can be accessed at:~\url{https://youtu.be/EOqlirEFGbg}.

\begin{figure}
\centering
\includegraphics[width=\linewidth]{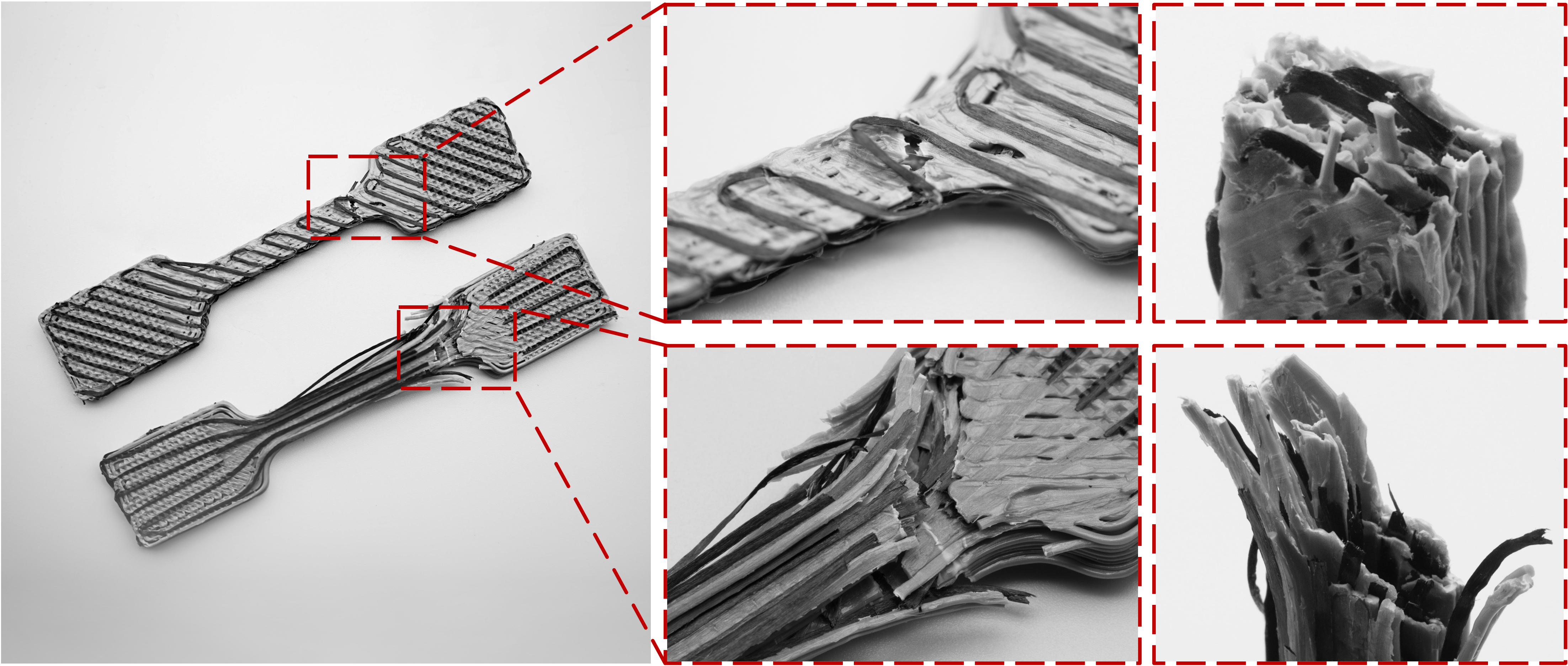}
\caption{The comparison of failure modes for CFRTPCs fabricated by using different toolpaths -- (top) the conventional zig-zag toolpaths and (bottom) our stress-oriented adaptive CCF toolpaths.}
\label{fig:FailureComparison}
\end{figure}

\begin{figure}[t]
\centering
\includegraphics[width=\linewidth]{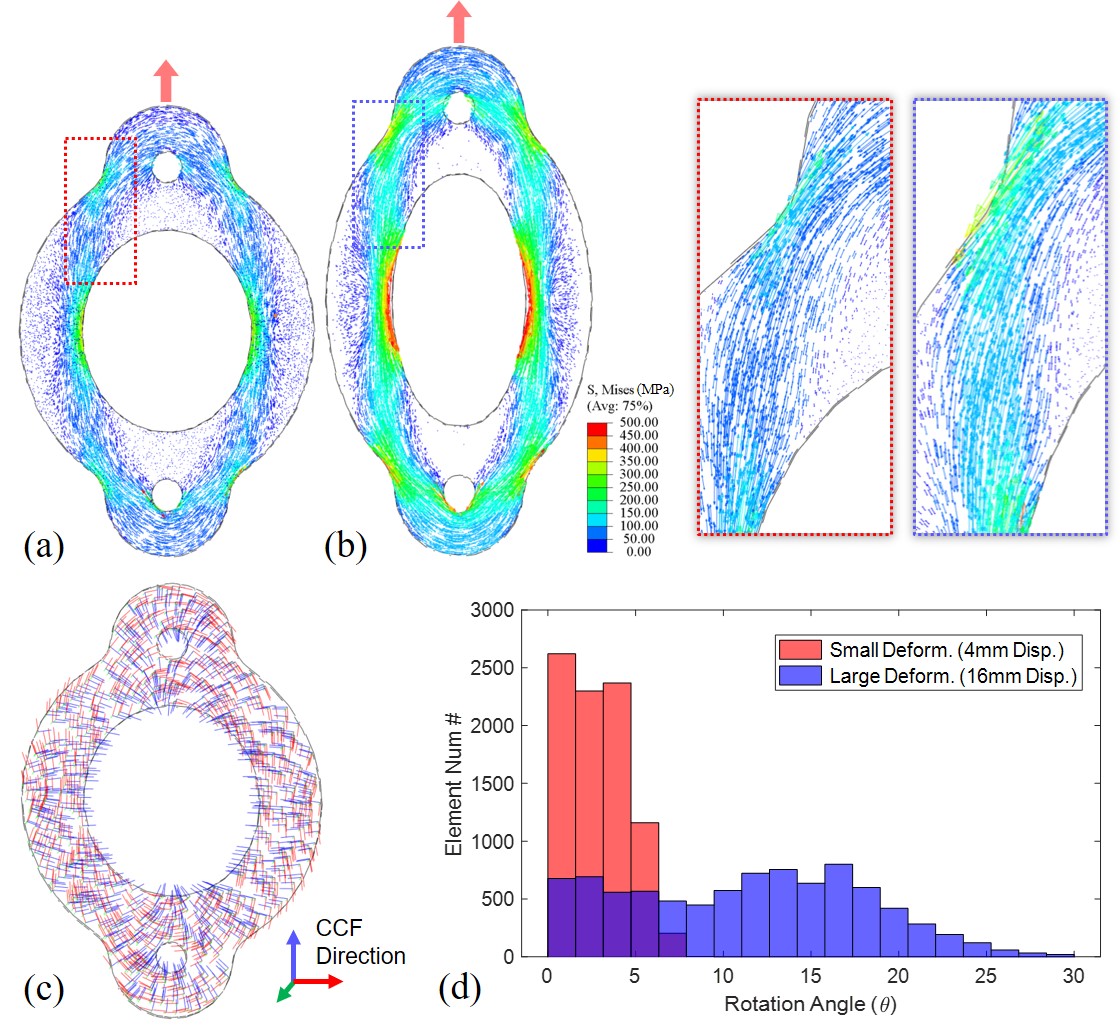}
\caption{Analysis of principal stresses' directional change for the Loop model by using FEM simulation with anisotropic material specified at the element level. (a) The vector field of maximal principal stresses under the load of applying 4 mm displacement. (b) The field of maximal principal stresses under the load of 16 mm displacement, where the directions of maximal principal stresses are changed in many regions (e.g., the region circled by dash lines). (c) Visualization of the local frames used to define the direction of CCF reinforcement in our method, which is also used to specify the local direction of anisotropic material in FEM simulation. (d) The histogram to show the rotations of maximal principal stresses in elements under different loads.
}
\label{fig:anisotropicSim}
\end{figure}

\subsection{Discussion}
There is a concern about whether using adaptive CCF toolpaths will result in weaker bonding between layers than zig-zag toolpaths. In our experimental tests, we did not observe this drawback. This is mainly because of two major reasons. The first reason is that the CCF layers are very sparse and thin (i.e., $0.05$mm in our tests) while the PLA layers have a thickness of $0.8$mm, and the chemical bonding between CCFs and PLA is in general very weak no matter which pattern of CCF toolpaths is employed. Secondly, the adaptive toolpaths are only applied to the layers of CCFs, and PLA filaments are still fabricated by zig-zag toolpaths along perpendicular directions between neighboring layers. Therefore, the material bonding between PLA layers is still very strong in regions without CCFs.

It is interesting to observe the different modes of material failure on CFRTPCs fabricated by toolpaths in different patterns. As shown in the top of Fig.~\ref{fig:FailureComparison}, the structure failure of CFRTPC specimen with zig-zag CCF toolpaths is mainly caused by the delamination between PLA layers. Therefore, complete CCF can be observed after structural failure. Differently, when aligning CCFs along the major stress directions by using our stress-oriented adaptive toolpath, structural failure occurs when the fibres are broken -- see the bottom row of Fig.~\ref{fig:FailureComparison}. In summary, the anisotropic mechanical strength of CCF is better utilized.

Although the mechanical strength of 3D printed models can be significantly reinforced by using our CCF toolpaths, the effectiveness of this loading-dependent reinforcement can be reduced when the reinforced model presents very large deformation. As shown in the supplementary video, the Shell model under large deformation can make the major tensile direction largely different from the principal stresses in its original shape. Therefore, the delamination occurs between the CCF and the resin matrix. 

We prove this hypothesis by FEM simulation using Abaqus, where anisotropic material is defined at the element level. The direction and norm value of $\mathbf{u}_f$ are used to locally define the anisotropic material property for an element $f$. The significant directional change of principal stresses under large deformation can be found from the results of simulation as shown in Fig.~\ref{fig:anisotropicSim} -- e.g., in the region highlighted by the boxes of dash lines. When the loop model is under large deformation such as 16 mm in displacement, which is more than $20\%$ of the model's initial length, the computed CCF toolpath is no longer well-aligned with the directions of maximal principal stresses. As illustrated in Fig.~\ref{fig:anisotropicSim}(d), more than half of the elements have the directions of their principal stresses rotated with angles greater than 10 degrees. This in fact leads to large transversal loads applied to the filaments of CCF -- i.e., the weaker directions of 3D printed CFRTPCs. As a result, the strength of CCF along axial direction cannot reinforce the printed models any more.
%
%

In our current approach, the contour-zigzag toolpaths are employed to fabricate the layers of resin matrix. Ideally, using stress-oriented toolpaths such as the method developed in \cite{Fang20_TOG} together with the adaptive CCF toolpaths introduced in this paper can further enhance the mechanical strength of 3D printed models. We indeed observed this phenomenon when applying this strategy on the tensile-bar model (i.e., Models A-E shown in Fig.\ref{fig:DiffPatternsTensileTest}). The breaking force can be further enhanced by $0.212\mathrm{kN}$ (Model A), $0.029\mathrm{kN}$ (Model B), $0.144\mathrm{kN}$ (Model C), $0.681\mathrm{kN}$ (Model D) and $0.640\mathrm{kN}$ (Model E). However, we observe from our experimental tests that the loading-independent toolpaths for resin matrix can help enhance the mechanical strength when large deformation occurs. Therefore, we do not apply the stress-aware toolpaths for resin matrix. How to improve the robustness of the stress-oriented CCF toolpaths under large deformation will be explored in our future work. Moreover, as discussed in \cite{Zhang21_AdditManu}, the problems of fibre misalignment and breakage during fibre deposition will also reduce the effectiveness of CCF reinforcement. How to generate toolpaths that can prevent these manufacturing failure cases is also possible future work. 

%

%

%% file: Text/Sec7Conclusion.tex
\section{Conclusion}
\label{secConclusion}
This paper presents a field-based method of toolpath generation for 3D printing CFRTPCs. Our method has fully considered the strong anisotropic material property of continuous fibres and the effectiveness of adaptive density controlled according to the values of stresses. Algorithms based on vector and scalar fields are presented in this paper to generate stress-oriented adaptive CCF toolpaths. The effectiveness of our approach has been verified on a variety of models in different loading conditions. According to the experimental tests on specimens of CFRTPCs by using PLA as matrix, improvement of mechanical strength up to 71.4\% can be observed compared to reinforcement by using loading-independent CCF toolpaths with the same amount of CCF filament.